\RequirePackage{amsmath}
\documentclass[12pt]{iopart}

\usepackage{graphicx}
\usepackage{epstopdf}
\usepackage{color}
\usepackage{appendix}
\usepackage{bbm}
\usepackage{iopams}   
\newcommand{\X}{~I}
\newcommand{\Y}{~II}

\begin{document}
\title[]{Work, heat and entropy production in bipartite quantum systems}

\author{Hoda Hossein-Nejad, Edward J O'Reilly, Alexandra Olaya-Castro}

\address{Department of Physics and Astronomy, University College London, Gower Street, London WC1E 6BT, UK}
\ead{a.olaya@ucl.ac.uk}
\begin{abstract}
In bipartite quantum systems commutation relations between the Hamiltonian of each subsystem and the interaction impose fundamental constraints on the dynamics of each partition. Here we investigate work, heat and entropy production in bipartite systems  characterized by particular commutators between their local Hamiltonians and the interaction operator.
 We consider the formalism of [Weimer, EPL, 83:30008, 2008], in which heat (work) is identified with energy changes that (do not) alter the local von Neumann entropy, as observed in an effective local measurement basis. 
 We demonstrate the consequences of the commutation relations on the work and heat fluxes into each partition, and extend the formalism to open quantum systems where one, or both, partitions are subject to a Markovian thermal bath. 
  We also discuss the relation between heat and entropy in bipartite quantum systems out of thermal equilibrium, and reconcile the aforementioned approach with the second law of thermodynamics.
\end{abstract}

 \section{Introduction }

The emergence of thermodynamic behaviour within quantum mechanical systems has attracted much attention in recent years \cite{Kosloff:2013, Jarzynski:2015}.  
This interest is motivated by the desire to gain a deeper understanding of thermodynamics, as well as the need to obtain generalizations of thermodynamic concepts and constraints to quantum settings. 
A conceptual understanding of work, heat and the second law in the quantum domain has therefore been central to research in quantum thermodynamics  \cite{Allahverdyan:2004, Quan:2007, Talkner:2007, Gelbwaser:2013, Salmilehto:2014, Brandao:2015}. Owing to advancements in quantum technologies, experimental tests of these concepts are within tangible reach \cite{Dorner:2013, CampisiJ:2013, Mazzola:2013, Koski:2014}. The developments in this area allow investigation of the relevance of coherence in the efficiency of quantum engines \cite{Scully:2011} and can guide the design of nano-mechanical devices with thermodynamic functionalities \cite{Birjukov:2008, Horodecki:2013}.

 The conventional definition of work and heat for a quantum system evolving under a time-dependent Hamiltonian considers the change in the internal energy $U$ of a system as $\dot{U} = \mbox{Tr} \{  \rho \dot{H} \} + \mbox{Tr} \{  \dot{\rho} H \}$ and identifies the first (second) term as the work (heat) flux \cite{Alicki:1979}.
This division assumes that a classical, external, driving gives rise to a time-dependent Hamiltonian. 
In contrast,  we are interested in autonomous bipartite quantum systems in which the inherent quantum mechanical interaction between the two partitions results in internal transfer of energy. This internal transfer can subsequently be identified as either a heat, or a work, flux  \cite{Weimer:2008, Schroeder:2010, Teifel:2011}. 
 There are plenty of quantum settings where such is the case, but an interesting example is a prototype  biological dimer where the interaction between electronic and intramolecular vibrational motions results in non-classical behaviour and internal energy transfer \cite{OReilly:2014}.
A framework for defining work and heat fluxes 
in autonomous quantum systems was
 presented by Weimer and colleagues in Ref.~\cite{Weimer:2008}. 
This approach, henceforth referred to as the Weimer framework or the energy flux formalism, argues that time-dependent Hamiltonians arise 
from tracing out the degrees of freedom of the driver and 
need not have  a classical, or external, origin. 
By defining an effective local energy basis, heat (work) flux is identified with energy changes that (do not) alter the the local von Neumann entropy.    
This framework is of considerable conceptual interest as it 
aims to extend the connection between entropy and heat to finite, out of equilibrium quantum systems.
Despite its conceptual ease and formal structure, this approach has not been widely used in the investigation of work and heat in quantum systems \cite{Gemmer:2009}. This is arguably due to the differences between the work flux defined in this way and  
the extractable (measurable) work; the latter being the focus of most prior research \cite{Mazzola:2013, Talkner:2007}.  
As work done on a system is not a quantum observable \cite{Talkner:2007}, its quantum statistical properties and the backaction of the measurement apparatus need to be considered within any formalism. These issues have partially been addressed within the energy flux formalism \cite{Teifel:2011} but remain under scrutiny.

The simplicity of the energy flux formalism, however, merits the investigation of aspects scarcely explored such as the implications of conservation laws on the nature of energy exchanges and how the framework can be generalized to open quantum system dynamics. Conserved quantities (defined by Hermitian operators that commute with the total Hamiltonian) and their associated symmetries are of broad interest in physics. Moreover, transformation of symmetry-imposed constraints in open quantum systems has gained renewed attention in recent years \cite{Marvian:2014}.
In this article, we discuss work and heat exchanges within bipartite systems characterized by particular commutation relations between the local Hamiltonians and the interaction operator. 
We show how conserved quantities associated to a given partition, that arise from particular commutation relations, give rise to an asymmetric structure for the work and heat fluxes of the subsystems.
We furthermore demonstrate  the consistency of the approach when one partition acquires the character of a macroscopic bath, and extend the formalism to situations where one, or both, partitions are subject to a Markovian thermal bath. The connection between heat, as defined in this framework, and entropy is explored in the last section. This connection allows us to discuss the second law of thermodynamics within the energy flux formalism and demonstrate the compatibility of the approach with the second law. The present work strengthens the energy flux formalism and provides further evidence that thermodynamic quantities such as work and heat can be generalized to quantum systems far from thermal equilibrium. 
\begin{figure}[t]
\centering
\includegraphics[width=80mm]{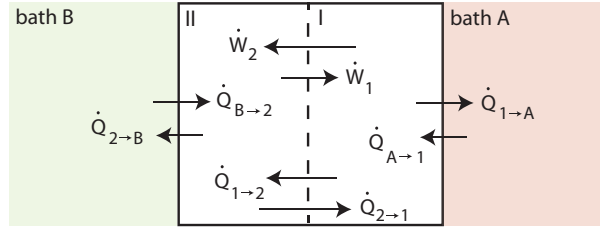}
\label{fig_2}
\caption{A schematic illustration of the bipartite quantum system investigated in this article. In section 2, we focus on the closed bipartite system (without a heat bath).
 In section 3, partition I is subject to a thermal bath.
  In section 4, both partitions interact with a thermal bath. 
In our notation the total heat flux into system I is written as $\dot{Q}_1 = \dot{Q}_{2 \rightarrow 1} + \dot{Q}_{\textrm{bath A} \rightarrow 1}$ where $\dot{Q}_{2 \rightarrow 1}$ and $\dot{Q}_{\textrm{bath A} \rightarrow 1}$
are the heat fluxes from system II and the bath A respectively. The work flux into system I is identified as $\dot{W}_1$.}
\end{figure}
 \section{Work and heat in closed bipartite quantum systems} 
\subsection{Review of the concepts} 
 In this section we briefly review the division  of the energy exchange between bipartitions of a quantum system into heat and work, as put forward by Ref.~\cite{Weimer:2008}. Within the framework open quantum systems, the total system is denoted as \textit{closed} when it  does not exchange energy or matter with an environment and is described by the Hamiltonian 
 $H = H_1 + H_2 + H_{12}$ . 
 Here  $H_1$ and $H_2$ are the individual Hamiltonians of subsystems I and II, and $H_{12}$ specifies the interaction between the two partitions. Notice that in classical thermodynamics our total system would be referred as \textit{isolated}. The density matrix of the total system can be written as
\begin{equation}
\rho (t) = \rho_1 (t) \otimes \rho_2(t) + C_{12} (t),
\label{corrole}
\end{equation}
where $\rho_1 (t)= \mbox{Tr}_2 \big\{ \rho (t) \big\}$,
 $\rho_2 (t)= \mbox{Tr}_1 \big\{ \rho (t) \big\}$ 
 and $C_{12} (t)$ represents the correlations between the 
 two partitions. We assume an initial separable state of the form
 $\rho (0) = \rho_1 (0) \otimes \rho_2 (0)$, with no bipartite correlations. 
 Starting from the equation of motion for the total system, and taking the trace with respect to system\Y, the Liouville-von Neumann equation for system\X~can be written as
\begin{equation}
\dot{\rho}_1 (t) = -i [H_1 + H_1^{\mbox{eff}} (t), \rho_1 (t)] -i  \mbox{Tr}_2 \big\{ [ H_{12}, C_{12} (t)] \big\},
\end{equation}
where the effective Hamiltonian of system\X~is identified to be 
\begin{equation}
H_1^{\mbox{eff}} (t) = \mbox{Tr}_2 \big\{ H_{12} [\mathbbm{1}_1   \otimes \rho_2 (t)] \big\}.
\end{equation}
The energy changes need to be defined with respect to a local 
effective measurement basis (LEMBAS), which is determined
 by the experimental apparatus. 
 This formalism assumes a LEMBAS that includes both the eigenstates 
 of the bare system Hamiltonian $H_1$, and the shift in the 
 eigenenergies of system I due to the interaction
  with system II. Since $H_1$ and $H_1^{\mbox{eff}}(t)$ may not be compatible observables at
  all times, the effective Hamiltonian is divided into two contributions
\begin{equation}
H_1^{\mbox{eff}} (t) = H_{1,a}^{\mbox{eff}} (t) + H_{1,b}^{\mbox{eff}} (t),
\label{eq:lembas}
\end{equation}
where $H_{1,a}^{\mbox{eff}} (t) $  and $H_{1,b}^{\mbox{eff}} (t) $ are components of the effective Hamiltonian that commute and do not commute with $H_1$ respectively. The commuting contribution of the interaction induces a shift in the local energy of system\X. The local basis is thus associated to measurements of the following operator
\begin{equation}
H_1'(t) = H_1 + H_{1,a}^{\mbox{eff}} (t),
\end{equation}
and the internal energy change of system I can be identified as
\begin{equation}
dU_1 =  \mbox{Tr}_1 \big\{ \rho_1  dH_1' (t)  \big\} + \mbox{Tr}_1 \big\{ H_1' (t) d \rho_1   \big\}. 
\label{div}
\end{equation}
We now divide the internal energy flux of system I into two contributions: the energy change that has no effect on the local von Neumann entropy, is labelled as the work flux $\dot{W}_1(t)$,  whereas the contribution that alters this entropy is identified as the heat flux $\dot{Q}_1(t)$.  This constitutes a statement of the first law of thermodynamics for partition I. The final
expressions read
\begin{equation}
\dot{W}_1 (t) = \mbox{Tr}_1 \big\{ 
\dot{H}^{\mbox{eff}}_{1,a} (t) \rho_1(t) - i [H_1' (t), H^{\mbox{eff}}_{1,b} (t)] \rho_1 (t)
\big\},
\label{work-mahler}
\end{equation}
\begin{equation}
\dot{Q}_1(t) = \mbox{Tr}_1 \big\{ H_1' (t) {\cal{L}}^{\mbox{eff}}_1 [\rho(t)]
\big\}.
\label{heat}
\end{equation}
where $
{\cal{L}}^{\mbox{eff}}_1 [\rho(t)] = -i  \mbox{Tr}_2 \big\{ [ H_{12}, C_{12} (t)] \big\}
$.
Note that at this point there is no assertion that work should be something more useful that heat \cite{Gemmer:2009}. Importantly, entropy changes can arise without an associated heat flux as is discussed in section 4.
Heat flux can, alternatively, be expressed as
\begin{eqnarray}
\dot{Q}_1(t)  =
    -i \mbox{Tr} \big\{ [ ( H_1' (t) \otimes \mathbbm{1}_2)  ,   H_{12} ]  C_{12} (t)  \big\}, 
\label{heating}
\end{eqnarray}
where the cyclic invariance of the trace has been used. 
Similar expressions can be written for system\Y~if measurements are performed in the 
 LEMBAS associated with $H_2$. 
 
Equations (\ref{work-mahler}) and (\ref{heating}) merit two important observations. 
Firstly, work flux now has an additional contribution 
due to the non-commuting part of the effective Hamiltonian in Eq. (\ref{eq:lembas}).
This is in contrast to the conventional definition that identifies work solely with the term $\mbox{Tr}_1 \big\{ \rho_1  dH_1' (t)  \big\}$ in Eq. (\ref{div}).
Secondly, the heat flux into system I vanishes if at least one of the 
following is satisfied: (i) the dynamics does not induce correlations between the two subsystems or  (ii) $H_1' (t)$ commutes
 with the interaction Hamiltonian. Thus, bipartite correlations are necessary, but insufficient, for the existence of a heat flux. We will discuss in section 4  the implications of this feature for the possible link between heat and entropy production in autonomous systems.
 
From the above discussion, it is clear that commutation relations between the different energy operators shape, in a non-trivial manner, the work and heat exchanges between the quantum partitions. In the following sections we consider three possible commutation relations between $H_1$, $H_2$ and $H_{12}$, and investigate their consequences 
on work and heat fluxes. Without loss of generality, we consider an interaction of the form $H_{12} = A_1 \otimes B_2$, where $A_1$ acts on system\X~and $B_2$ on system\Y. The effective Hamiltonian of system\X~is thus given by $H_1^{\mbox{eff}} (t) = A_1 \langle B_2 \rangle$.
All results can readily be extended to the more general interaction $H_{12} = \sum_i A_i \otimes B_i$. 
 \\
 \subsection{Fully commuting interaction}
A fully commuting Hamiltonian satisfies the relation: $ [A_1, H_1] = [B_2, H_2] = 0$. 
Since $[H_1^{\mbox{eff}} (t), H_1]=0$, we identify $H_{1,a} = H_1^{\mbox{eff}} (t)  $, and $H_{1,b}^{\mbox{eff}}  =0$. 
Work on system\X~is thus given by
\begin{eqnarray}
\dot{W}_1 (t)  =   \mbox{Tr} \big\{ 
\dot{H}^{\mbox{eff}}_{1,a} (t) \rho_1(t) 
\big\}   = 
\langle A_1 \rangle  \langle \dot{B}_2 \rangle
 = 0, 
\end{eqnarray}
where we have used $\langle  \dot{M} \rangle = -i \langle[M, H] \rangle$ for $M = \mathbbm{1}_1 \otimes B_2$. Similarly, one can show that $\dot{W}_{2}=0$.  From Eq. (\ref{heating}) it can be shown that the exchanged heat also vanishes, $\dot{Q}_1 (t) = \dot{Q}_2 (t) = 0 $. The energy exchange is thus zero, as expected for a commuting interaction.
An example of such a system is provided by the spin-oscillator Hamiltonian with a dispersive interaction: $H = \frac{\omega_0}{2} \sigma_z + \nu a^{\dagger} a + g  \sigma_z a^{\dagger} a $, where $\sigma_z$ is the population difference operator, $\omega_0$ is the energy gap between the ground and excited state of the spin, $\{a, a^{\dagger} \} $ are the annihilation and creation operators of the oscillator, $\nu$ is the oscillator frequency and $g$ quantifies the magnitude of the spin-oscillator coupling. 
\subsection{Partially commuting interaction}
We define a partiality commuting interaction via the commutator properties: $[H_{1}, A_1] = 0$ and $[H_{2}, B_2] \neq 0$ i.e. observable $A_1$ is conserved. 
Since $[H_1^{\mbox{eff}} (t), H_1]=0$, we identify $H_{1,a}^{\mbox{eff}} (t) = H_1^{\mbox{eff}} (t)  $, and $H_{1,b}^{\mbox{eff}}  =0$. Work on system\X~is thus given by
\begin{eqnarray}
\dot{W}_1 (t) =  \mbox{Tr}_1 \big\{ 
\dot{H}^{\mbox{eff}}_{1,a} (t) \rho_1(t) 
\big\} = 
\langle A_1 \rangle  \langle \dot{B}_2 \rangle. 
\end{eqnarray}

Work done on system\Y~can be computed in a similar manner. The effective Hamiltonian of system\Y~is given by
$
H_2^{\mbox{eff}} (t) = \langle A_1 \rangle  B_2
$.
Unlike the previous case, $H_2^{\mbox{eff}} (t)$ has no commuting contributions with $H_2$. That is,  $H_{2,a}^{\mbox{eff}}  = 0 $, and $H_{1,b}^{\mbox{eff}} = H_2^{\mbox{eff}} (t)$. 
The work done on system\Y~is thus given by
\begin{eqnarray}
\dot{W}_2 (t) 
&  = &  -i \mbox{Tr}_2 \big\{ [ H_2 ,  \langle A_1 \rangle  B_2 ] \rho_2 (t) \big \}, \nonumber \\
& = &    -i \mbox{Tr}_{1,2}  \big\{ [  \mathbbm{1}_1 \otimes H_2 ,   A_1 \otimes  B_2 ] \rho_1 (t) \otimes \rho_2 (t)  \big\},
\nonumber
\\
&  = &-  \langle A_1 \rangle  \langle \dot{B}_2 \rangle .
\end{eqnarray}
The general work flux relation holding for any partially commuting interaction is thus
\begin{equation}
\dot{W}_1 (t) + \dot{W}_2 (t) =0.
\label{eq:pc_closed}
\end{equation}
The heat fluxes into systems I and II are computed  from Eq. (\ref{heating}), leading to 
\begin{eqnarray}
\dot{Q}_1(t) & = & 0,  \\ 
  \dot{Q}_2(t) & \neq &  0, 
\end{eqnarray}
since 
$[ ( H_1' (t) \otimes \mathbbm{1}_2 )  ,   H_{12} ]= 0$  while  $[ ( \mathbbm{1}_1 \otimes  H_2' (t)  )  ,   H_{12} ] \neq 0$.
Note that control of the system Hamiltonian without introducing heat is reminiscent of classical driving. System II thus acts as a classical drive for system I, while the converse is not true.
Moreover, for this commutation scenario, heat flux is exclusively from system I to system II, and the existence of correlations between the two partitions determines whether or not system\Y~receives heat. This observation is a direct consequence of conservation of the observable $A_1$, and does not violate the conservation of energy.

Under what circumstances do the correlations, and therefore $\dot{Q}_2(t)$, vanish? Note that the commutator $[H_1, A_1] = 0$, ensures that $H_1$ and $A_1$ are simultaneously diagonalizable. $A_1$ can thus be expanded in the energy eigenbasis of $H_1$. Denoting the eigenstates of $H_1$ by $\{ |i \rangle \}$, $A_1$ can be written as
$A_1 = \sum_i a_i |i \rangle \langle i |$, 
where $\{a_i \}$ are a set of coefficients. The interaction Hamiltonian is thus given by 
\begin{equation}
H_{12} = \sum_i a_i |i \rangle \langle i | \otimes B_2.
\label{int}
\end{equation}
From this Hamiltonian one can construct an evolution operator and determine the dynamics of the density matrix $\rho(t)$ for an initially separable state $\rho (0) = \rho_1 (0) \otimes \rho_2(0) $. If the initial state of system\X~is an incoherent mixture of eigenstates $\rho_1 (0)  =\sum_i c_i  | i \rangle \langle i |$, the dynamics are given by
\begin{equation}
\rho(t) =\sum_i c_i |i \rangle \langle i |\otimes e^{-i(H_2 + a_i B_2) t} \rho_2(0)  e^{i(H_2 + a_i B_2) t}. 
\end{equation}
The total state only remains separable if all but one of the coefficients $c_i$, are zero. 
i.e.:when system I starts in an energy eigenstate of $H_1$. We conclude that the dynamics do not correlate the two subsystems, i.e. $\dot{Q}_2(t)$=0, if and only if system I is initialized in an eigenstate.

An example of such a quantum setting is  the displaced spin-oscillator Hamiltonain:  $H = \frac{\omega_0}{2} \sigma_z + \nu a^{\dagger} a + g  \sigma_z (a^{\dagger} + a) $, where $H_1 =  \frac{\omega_0}{2} \sigma_z $ and  $H_2 = \nu a^{\dagger} a  $. Denoting the eigenstates of the spin by $ \{ |i \rangle \}$ where $ |i \rangle = \{ |e \rangle, |g \rangle \} $, for the initial state $\rho (0) = |i\rangle \langle i| \otimes \rho_2 (0)$, the spin and the oscillator evolve independently and no heat is being transferred to the oscillator. However, if the spin is initially prepared in a mixed state of the form $\rho_1(0) = c |g \rangle \langle g |+ (1-c) |e \rangle \langle e | $, the dynamics correlate the two subsystems, generating a heat flux into the oscillator. 
To illustrate the latter point, let us assume that the oscillator is initially prepared in a coherent state, while the spin is prepared in a mixture of eigenstates. This example was considered in Ref. \cite{Teifel:2011} and the authors  conclude that there is no build up of correlations between the spin and the oscillator.
 In the remaining part of this section, we demonstrate that this conclusion is incorrect; 
 although the heat flux into the spin is zero ($\dot{Q}_1 = 0$), there is finite heat flux into the oscillator ($\dot{Q}_2 \neq 0$) for an initial mixed state of the spin. The state of the spin and the oscillator may therefore become (classically) correlated. 
For this initial state, the work flux into the spin can be shown to be 
\begin{equation}
\dot{W}_1 (t)  = g (1-2c) \langle \dot{x} (t) \rangle , 
\end{equation}
where $x = a+ a^{\dagger}$. We assume an initial coherent state denoted by $|\alpha_0 \rangle$ with displacement $x_0 \equiv \mbox{Re} [\alpha_0]  $ and zero initial momentum, $p_0 \equiv\mbox{Im}[\alpha_0 ] = 0$. The evolution of the density matrix of the oscillator is given by
\begin{equation}
\rho (t)= c | \alpha_{-} (t) \rangle  \langle \alpha_{-} (t) |  +  (1-c)  | \alpha_{+} (t) \rangle \langle  \alpha_{+} (t) |, 
\end{equation}
where $|\alpha_{+} (t)\rangle $ and $|\alpha_{-} (t)\rangle $ are the displaced coherent states: 
\begin{equation}
|\alpha_{\pm} (t) \rangle  = \exp \bigg\{  {-i \bigg( \nu a^{\dagger} a \pm g (a + a^{\dagger}) } \bigg) t \bigg\} |\alpha_0 \rangle .
\end{equation}
The expectation value of the position of the oscillator is given by 
\begin{equation}
\langle x (t) \rangle = c \langle x \rangle_{\alpha_{-} (t)}   +  (1-c)  \langle x \rangle_{\alpha_{+} (t)}.
\end{equation}
The expectation value of the oscillator velocity is the quantity of interest for evaluation of the work flux and can be shown to be  
\begin{equation}
\langle \dot{x}  (t)  \rangle_{{\alpha_\pm (t)} } = 
2 \nu
\sin (\nu t)
\bigg( x_0 \pm \frac{g}{\nu} \bigg), 
\end{equation}
where the coupling $g$ is also assumed to be real. The final expression for the work flux into the spin thus reads
\begin{equation}
\dot{W}_1 (t) = 2 g  (1-2c)
\big[ \nu x_0 + (1-2c) g \big]
\sin( \nu t). 
\end{equation}
The corresponding heat fluxes are found to be, 
\begin{eqnarray}
\dot{Q}_1  (t) & = &  0, \\   
\dot{Q}_2  (t) & =&  8 c(1-c) g^2 \sin(\nu t). 
\end{eqnarray} 
 The heat flux into the oscillator is independent of the initial displacement of the oscillator and is maximum for a maximally mixed initial state of the spin ($c=1/2$). The work flux, however, is  maximized for an initially pure state ($c=1$ or $c=0$).  The initial state of the spin thus serves as a control parameter that  will determine the nature of energy exchange between the two subsystems. 

\subsection{Non-commuting interaction}
This scenario is characterized by the commutators: $[A_1, H_1] \neq 0$, and $[B_2, H_2] \neq 0$. From the effective Hamiltonian of system\X, we identify $H_{1,a}^{\mbox{eff}} (t) = 0 $, and $H_{1,b}^{\mbox{eff}} (t) = H_1^{\mbox{eff}} (t) $. Work on system\X~is thus given by
\begin{eqnarray}
\dot{W}_1 (t)   =   -i \mbox{Tr}_1 \big\{ [ H_1 , H^{\mbox{eff}}_{1,b} (t) ] \rho_1 (t)
\big\}, 
\end{eqnarray}
and similarly for system\Y: 
$
\dot{W}_2 (t) =   -i \mbox{Tr}_2 \big\{ [ H_2 , H^{\mbox{eff}}_{2,b} (t) ] \rho_2 (t)
\big\} . 
$
Heat flux into system\X~is given by
\begin{eqnarray}
\dot{Q}_1(t) 
  & = & 
    -i \mbox{Tr} \big\{ [ ( H_1  \otimes \mathbbm{1}_2 )  ,   H_{12} ]  C_{12} (t)   \big\}. 
\end{eqnarray}
Similar expressions can be obtained  for system\Y.  
The heat fluxes $\dot{Q}_{1} (t)$ and $\dot{Q}_{2} (t)$ are, in general, non-vanishing and unequal. As the dynamics always correlate the two subsystems, regardless of the choice of the initial state, $C_{12} (t)$ cannot be zero for all $t$. If the condition $[H_1, H_{12}] = - [H_2, H_{12}]$ is satisfied, however, the net heat and work fluxes will be zero, that is $\dot{Q}_1 (t)  + \dot{Q}_2 (t)  = 0$ and $\dot{W}_1 (t)  + \dot{W}_2 (t)  = 0$. 

The Jaynes Cummings model with the rotating wave approximation is an example of such an interaction: $H = \frac{\omega_0}{2} \sigma_z + \nu a^{\dagger} a + g  (\sigma_{+} a + \sigma_{-} a^{\dagger})$ where $\sigma_{+}$ and $\sigma_{-}$ are the raising and lowering operators of the spin. The condition $[H_1, H_{12}] = - [H_2, H_{12}]$ translates to an oscillator that is on resonance with the spin, $\nu = \omega_0$.   

\section{Generalization to open systems}
\subsection{One partition becomes a Markovian bath}
In this section  we demonstrate that heat flux into system I converges towards known results, in the limit that system\Y~acquires the character of a macroscopic Markovian thermal  bath.  The exact equation of motion for $\rho_1 (t)$ is given by
\begin{equation}
\dot{\rho}_1 (t) = - i [H_1,\rho_1 (t) ] - i \mbox{Tr}_2[H_{12}, \rho(t)].
\end{equation}
Recalling that $\rho(t) = \rho_1 (t) \otimes \rho_2 (t) + C_{12} (t)$, we obtain
\begin{equation}
\dot{\rho}_1 (t) = - i [H_1,\rho_1 (t) ] - i \mbox{Tr}_2[H_{12}, \rho_1 (t) \otimes \rho_2 (t) + C_{12} (t)].
\label{cool_dude}
\end{equation}
As system II is now a thermal bath,  we make the substitution $\rho_2 (t) \rightarrow \rho_B (t)$ and assume a general system-bath interaction of the form $H_{1B} = X_1 \otimes Y_B$, such that
\begin{equation}
\dot{\rho}_1 (t) = - i [H_1 + X_1 \langle Y_B \rangle, 
\rho_1 (t) ] - i \mbox{Tr}_B[H_{1B}, C_{1B} (t)].
\label{correctington}
\end{equation}
For weak system-bath interactions, the expectation value $\langle Y_B \rangle$ can be approximated by its initial  thermal value. The term $X_1 \langle Y_ B\rangle$ thus represents a bath-induced renormalization of the system energy.
For brevity we assume $\langle Y_B \rangle  = 0$.
Bath operators are often defined to have vanishing fluctuations such that the influence of the bath can be entirely characterized through a two-time correlation function \cite{Breuer:2002}. This is true for a thermal bath, provided that $[H_B, Y_B] \neq 0$.  
The case of constant $\langle Y_B \rangle$  renormalizes the system Hamiltonian and should be accounted for in the derivations that follow.
Consistency demands that the second term in Eq. (\ref{correctington}) approaches a Lindblad dissipator in the Markov limit: 
\begin{equation}
-i \mbox{Tr}_B[H_{1B}, C_{1B} (t)]\xrightarrow[]{\textrm{Born-Markov }}
{\mathcal{L}}_1 [\rho_1(t)].
\label{limit_ing}
\end{equation} 
This indicates that, even in the Markov limit, the existence of system-environment correlations is crucial for  transfer of heat. The heat flux then takes the expected form in this limit, i.e. $\dot Q_1(t)=\mbox{Tr}_1 \big \{ H_1 {\mathcal{L}}_1 [\rho_1(t)]\big \}$.

 Eq. (\ref{limit_ing}) implies that system-bath correlations are not absent in the Markov limit. Note that there is no contradiction between this observation and the Born approximation used in the derivation of Markovian master equation, which assumes separability of the state of the system and the bath, $\rho \approx \rho_s (t) \otimes \rho_b$. 
 Indeed, the Born approximation does not imply that there are no excitations in the bath caused by the system \cite{Breuer:2002}. 
   The combination of the Born and the Markov approximations provide a description on a coarse-grained time scale.   The assumption is that environmental excitations decay on a time scale that is much faster than the time scale of the system dynamics, and thus cannot be resolved. Over this coarse-grained time scale, the state of the bath can be assumed to be approximately constant, and the system-bath correlations can be neglected. In actuality, build-up of correlations between the system and the bath leads to loss of phase information from the system. Care therefore must be taken in applying the Born approximation outside the scope of the derivation of the Markovian master equation. 

\subsection{One partition interacts with a Markovian bath}
The notion of a quantum system that serves as a time-dependent drive for a second quantum system, can still be applied if either, or both, partitions interact with a bath. In this section we present this generalization to open quantum systems. Again, note here the semantic difference with classical thermodynamics where such system would be denoted as `closed' indicating that it may exchange energy (but not matter) with a bath. 
Consider an open bipartite system, where only one partition (always I) interacts with a Markovian bath. The Hamiltonian of the total system, including the bath, can be written as
\begin{equation}
H = H_1 + H_2 + H_{12} + H_B + H_{1B}, 
\end{equation}
where $H_B$ is the bath Hamiltonian and $H_{1 B}$ describes the interaction of the bath with partition I. The action of the bath is described by a Markovian dissipator in the Lindblad form ${\mathcal{L} }_1[\rho_1]$, and the equations of motion for $\rho_1 (t)$ and $\rho_2 (t)$ now read
\begin{equation}
 \dot{\rho_1} (t)  =   -i [H_1 + H_1^{\mbox{eff}} (t), \rho_1 (t)] + {\cal{L}} ^{\mbox{eff}}_1 [\rho(t)] + {\cal{L} }_1[\rho_1],
 \end{equation}
 \begin{equation}
 \dot{\rho_2} (t)  =   -i [H_2 + H_2^{\mbox{eff}} (t), \rho_2 (t)] + {\cal{L}}^{\mbox{eff}}_2  [\rho(t)].
\end{equation}
The heat flux to system\X~now has an extra contribution due to the bath: 
\begin{eqnarray}
\small{
 \dot{Q}_1 (t)  =  \mbox{Tr}_1 \big\{ H_1'(t) {\cal{L}}_1^{\mbox{eff}} [\rho(t)]
 \big\} + 
 \mbox{Tr}_1 \big\{
 H_1'(t) {\cal{L}}_1 [\rho_1(t)]
 \big\}, \label{heat_up} 
 }
 \end{eqnarray}
 \begin{equation}
 \dot{Q}_2 (t)  =  \mbox{Tr}_2 \big\{ H_2'(t) {\cal{L}}_2^{\mbox{eff}} [\rho(t)]
 \big\}. 
\end{equation}
We consider the case where  $[H_1' (t), H_{1B}] \neq 0$, such that the bath can exchange energy with system I, and  investigate the modification to the work and heat flux relations derived in section 2. We will not discuss the heat flux from system I to the bath, but note that, unless for special cases, its magnitude cannot easily be deduced from $\dot{Q}_1 (t)$. 

\subsubsection{A fully commuting interaction: $[H_{i}, H_{12}] = 0$, $i=\{ 1, 2\}$.} Unlike the corresponding closed system scenario (Sec. 2.2), the  work and heat fluxes on the subsystems are unequal prior to equilibration and are given by
\begin{equation}
 \dot{W}_1 (t) =   \langle \dot{B}_2 (t) \rangle  \langle A_1 (t) \rangle  = 0, 
\end{equation}
\begin{equation}
 \dot{W}_2 (t) =   \langle B_2 (t) \rangle  \langle \dot{A}_1 (t) \rangle 
 =
   \langle B_2 (t) \rangle
\mbox{Tr}_1  \big\{ A_1 {\cal{L}}_1 [\rho_1(t)]  \big\}.
\end{equation} 
Interestingly, the bath-induced relaxation of the operator $\langle A_1 (t) \rangle$ is manifested as a work flux on system II until  $\langle \dot{A}_1(t)\rangle=0$ at thermal equilibrium.
As for heat we obtain
\begin{eqnarray}
 \dot{Q}_1 (t) & = & \mbox{Tr}_1 \big\{
 H_1'(t) {\cal{L}}_1 [\rho_1(t)]
 \big\}, \\
 \dot{Q}_2 (t) & = & 0. 
\end{eqnarray}
The heat flux $\dot{Q}_1 (t)$, is maintained until system I is thermalized, at which point $ \dot{Q}_1 (t)  =  \dot{Q}_2 (t)  = 0$. 
 The dispersive spin-oscillator Hamiltonian with the spin exposed to a heat bath provides an example of this scenario: 
$H = \nu a^{\dagger} a + \omega_0 \sigma_z + g
 a^{\dagger} a  \sigma_z + H_B + H_{1\textrm{B}}$, where $H_B = \sum_k \omega_k b_k^{\dagger} b_k$, and  $H_{1\textrm{B}} = \sigma_x \otimes  \sum_k  \lambda_k    ( b_k +  b_k^{\dagger} )$. The interaction of the spin with the bath will result in (i) a net transfer of heat to the spin and (ii) coherent energy transfer from the spin to the oscillator. For instance, the oscillator can represent the mechanical mode of a resonator, the spin a two-level atom, and the bath the background radiation. The relaxation of the atom is subsequently manifested as a net work flux on the resonator.

\subsubsection{A partially commuting interaction.}
Given that in our analysis the bath always interacts with system\X, we consider two different scenarios of a partially commuting interaction to cover all possibilities. 

\textit{Case 1. $[H_1, H_{12}] = 0, [H_2, H_{12}] \neq 0$}. Work fluxes in this case are given by
\begin{equation}
\dot{W}_1 (t) = - \dot{W}_2 (t) = \langle A_1 \rangle   \langle \dot{B}_2 \rangle.
\label{work_flux_open}
\end{equation}
The condition $W_1 (t) + W_2 (t) = 0$ is thus still satisfied (cf. Eq. (\ref{eq:pc_closed})) despite the addition of the thermal bath. The heat fluxes are modified as follows, 
\begin{eqnarray}
 \dot{Q}_1 (t) & = & \mbox{Tr}_1 \big\{
 H_1'(t) {\cal{L}}_1 [\rho_1(t)]
 \big\}, \\
\dot{Q}_2(t) 
  & = & 
    -i \mbox{Tr} \big\{ [ (  \mathbbm{1}_1 \otimes H_2' (t)   )  ,   H_{12} ]  C_{12}  (t) \big\}. 
    \label{heat_osc}
\end{eqnarray}
In the corresponding closed scenarios, we observed that heat flux is solely into system\Y. This observation is no longer valid for the open system prior to thermal equilibrium. 

An example of such an interaction is provided by the displaced spin-oscillator Hamiltonian with the spin exposed to a bosonic bath: $H = \frac{\omega_0}{2} \sigma_z +
\nu a^{\dagger} a + g  \sigma_z (a^{\dagger} + a) + H_B + H_{1\textrm{B}}$, where $H_{1\textrm{B}} = \sigma_x \otimes \sum_k \lambda_k   ( b_k +  b_k^{\dagger} )$. 

\textit{Case 2. $[H_1, H_{12}] \neq 0, [H_2, H_{12}] = 0$}.
The work flux takes the same form as the previous case, but the heat flux is now exclusively into system\X: 
\begin{eqnarray}
  \dot{Q}_1 (t)  & = &    -i \mbox{Tr}  \big\{ [ (  \mathbbm{1}_1  \otimes H_2' (t)  )  ,   H_{12} ]  C_{12} (t)  \big\} 
   +  \mbox{Tr}_1 \big\{
 H_1'(t) {\cal{L}}_1 [\rho_1(t)]
 \big\},  \\
\dot{Q}_2(t) 
&   = &  0
   \end{eqnarray}
An example of such a system is provided by the spin-oscillator model where now only the oscillator interacts with a bosonic bath, 
that is,  $H = \frac{\omega_0}{2} \sigma_z + \nu a^{\dagger} a + g  \sigma_z (a^{\dagger} + a) + H_B + H_{1B} $, for $H_{1\textrm{B}} = (a + a^{\dagger}) \otimes  \sum_k  \lambda_k    ( b_k +  b_k^{\dagger} )$. 
\subsubsection{ A non-commuting interaction.}
For a fully non-commuting interaction, work flux associated to each subsystem is given by  
$\dot{W}_j (t)  =  -i \mbox{Tr}_j \{  [H_j, H^{\mbox{eff}}_{j,b} (t)   ] \rho_j  (t)   \} $.   
The heat fluxes take the form
\begin{eqnarray}
\dot{Q}_1(t) 
  & = & 
    -i \mbox{Tr} \big\{ [ ( H_1  \otimes \mathbbm{1}_2))  ,   H_{12} ]  C_{12} (t)   \big\}  
    +   \mbox{Tr}_1 \bigg\{
 H_1'(t) {\cal{L}}_1 [\rho_1(t)]
 \bigg\},\\
\dot{Q}_2(t) 
  & = & 
    -i \mbox{Tr} \big\{ [ (    \mathbbm{1}_1 \otimes H_2)  ,   H_{12} ]  C_{12} (t)  \big\}. 
\end{eqnarray}
Similar to the finite system case (Sec. 2.4.), if $[H_1, H_{12}] = - [H_2, H_{12}]$, then $\dot{W}_1 (t) + \dot{W} (t)=
0 $. At thermal equilibrium, where the heat flux due to the bath vanishes, this condition also implies that
$\dot{Q}_1 (t) + \dot{Q} (t)= 0$.  

 The Jaynes Cummings model is an example of such an interaction: $H = \frac{\omega_0}{2} \sigma_z + \nu a^{\dagger} a + g  (\sigma_{+} a + \sigma_{-} a^{\dagger}) + H_B + H_{1B}$. The
 bosonic bath can interact with either the spin [e.g.: $H_{1\textrm{B}} = \sigma_x \otimes \sum_k  \lambda_k  ( b_k + b_k^{\dagger} )$] or the oscillator [e.g.: $H_{1\textrm{B}} = (a+a^{\dagger}) \otimes \sum_k  \lambda_k    (b_k^{\dagger} + b_k^{\dagger} )$].

 \subsection{Each subsystem interacts with a bath}
 The present treatment can readily be extended to the case where both sub-systems interact with a thermal bath. We will only consider the case of the fully-commuting interaction for brevity, but the theory can straightforwardly be applied to all cases.  Consider a bi-partite system where each partition interacts with its own Markovian bath. The total Hamiltonian is given by
 \begin{equation}
 H = H_1 + H_2 + H_{12} + H_{B_1} + H_{B_2} + H_{1B} + H_{2B},
 \end{equation}
 where $H_{B_1}$ and $ H_{B_2} $ are the bath Hamiltonians, while $H_{1B}$ and $ H_{2B}$ specify the system-bath interactions. 
 The action of the two baths is described by the Markovian dissipators ${\cal{L} }_1[\rho_1]$ and ${\cal{L} }_2[\rho_2]$. The equations of motion for the partitions now take the form
 \begin{equation}
 \dot{\rho_j} (t)  =   -i [H_j + H_j^{\mbox{eff}} (t), \rho_j (t)] + {\cal{L}} ^{\mbox{eff}}_j [\rho(t)] + {\cal{L} }_j[\rho_j]. 
 \end{equation}
For the fully commuting interaction, work fluxes are given, 
$
\dot{W}_1 =   \langle \dot{B}_2 (t) \rangle \langle A_1 (t) \rangle, 
\dot{W}_2 =   \langle B_2 (t) \rangle \langle \dot{A}_1 (t) \rangle,
$
while the heat flux takes the form
$
 \dot{Q}_j (t) =  \mbox{Tr}_j \big\{
 H_j'(t) {\cal{L}}_j [\rho_j(t)]
 \big\}. 
$
In the steady state, energy conservation and vanishing work fluxes mean that all energy fluxes are zero.  For the fully non-commuting interaction, however, it can be shown that a finite heat and work flux can be maintained in the steady-state.
Even with a fully commuting interaction, more elaborate designs consisting of multiple baths of
different temperatures can be utilized to engineer thermodynamic
devices capable of maintaining a constant heat flux in the steady state. For instance, there are prototype
thermoelectric devices consisting of two interacting quantum dots coupled to
 three thermal reservoirs \cite{Sanchez:2011}. The coupling between the quantum dots is a Coulombic repulsion and therefore a fully-commuting interaction. It has been demonstrated both experimentally \cite{Hartmann:2015}
 and theoretically \cite{Sanchez:2011} that this design aids the extraction of energy from a
 temperature difference while avoiding any direct connection between the hot and cold  reservoirs along which heat might flow.

 \section{Entropy change and the second law} 
 In Ref. \cite{Weimer:2008} it is argued that entropy and heat flux in a finite quantum system are related via the expression 
 \begin{equation}
 dS_1 = \beta^* dQ_1, 
 \label{useless_s}
 \end{equation}
 where $d S_1$ is the change in the  von Neumann entropy of system I and  $\beta^*$ is defined to be an effective inverse temperature. This definition is inspired by analogy with classical thermodynamics and was used to demonstrate agreement with Clausius relation for systems at thermal equilibrium.
There is, however, a problem with this identification as counterexamples to Eq.~(\ref{useless_s}) may be found where the heat flux is vanishing ($dQ_1 = 0 $), but the entropy flux is finite ($ dS_1 \neq 0 $). To illustrate this point, consider the scenario discussed in section 2.2 where both subsystem Hamiltonians commute with the interaction. In this case, there is no heat flux into either of the partitions, yet the local entropy of system I can still change, provided that it has an initial coherence and that partition II does not start in an energy eigenstate. In appendix A we show that if these two conditions are satisfied, the evolution of partition I is non-unitary, and its local entropy is oscillatory, despite the lack of a heat flux, and in contradiction with Eq. (\ref{useless_s}).

Clearly, a different line of thought ought to be followed to reconcile the energy flux formalism with the second law. 
For an initial separable state $\rho(0) = \rho_1 (0) \otimes \rho_2 (0)$, evolving under the Hamiltonian $H =  H_1 + H_2 + H_{12}$,  Esposito and et al. \cite{Esposito:2010} have demonstrated that the change in the von Neuman entropy of system I, between $t=0$ and $t=\tau$, can be expressed as
 \begin{eqnarray}
 \Delta S_1(\tau)  & = &  S_1(\tau) - S_1 (0)  \nonumber
 \\ & = & D \bigg[ \rho(\tau) || \rho_{1} (\tau) \rho_2 (0) \bigg] +
  \mbox{Tr} \bigg\{  \big[ \rho_2 (\tau) - \rho_2 (0) \big]   \mbox{ln} \rho_2 (0)
  \bigg\} ~,
  \label{espo}
 \end{eqnarray}
where $D[\rho_1 || \rho_2 ] = \mbox{Tr} \{ \rho_1 \ln \rho_1 \} -  \mbox{Tr} \{ \rho_1 \ln \rho_2 \}$ is the relative entropy. This expression is derived from the conservation of the total entropy with the sole assumption of the separability of the initial states. The first term in Eq.~(\ref{espo}) is the irreversible contribution to the entropy change and is known as the \textit{entropy production}. The second term is the reversible contribution to the entropy change. It quantifies the exchange of heat and results from the reduced dynamical evolution of system II. The entropy change can therefore be written in the standard thermodynamic form $\Delta S_1 (\tau) = \Delta S_1^{\textrm{ir}} +  \Delta S_1^{\textrm{re}}$ with 
\begin{eqnarray}
\Delta S_1^{\textrm{ir}} (\tau) & = & D \bigg[\rho(\tau) || \rho_1 (\tau) \rho_2 (0)   \bigg]  \label{p0},\\
 \Delta S_1^{\textrm{re}} (\tau) & = &   \mbox{Tr} \bigg\{  \big[ \rho_2 (\tau) - \rho_2 (0) \big]   \mbox{ln} \rho_2 (0)
  \bigg\} \label{p1}. 
\end{eqnarray}
Since the relative entropy is positive, Eq. (\ref{p0}) indicates the positivity of the entropy production, i.e.: $\Delta S_1^{\textrm{ir}} (\tau) \ge 0$, thus fulfilling the second law of thermodynamics \cite{Breuer:2002, Spohn:1978}. 

Equations (\ref{p0}) and (\ref{p1}) can also be used to investigate any possible links between heat and entropy. 
Clearly, Eq.~(\ref{p1}) cannot be identified with a heat flux. In classical thermodynamics the association between heat and entropy holds for reversible entropy exchanges and is defined for a thermal initial state.  Eq. (\ref{p0}) shows that this association is not correct for the general quantum setting.
The irreversible entropy is due to correlations between the two partitions that may arise independently of a heat flux.  We therefore conclude that the direct association between heat flux and entropy change, as suggested by Eq.~(\ref{useless_s}), does not hold in general.
\\

As an example, equations (\ref{p0}) and (\ref{p1}) can be used to analyze the entropy flux for the commuting interactions (sec. 2.2.).
From Eq.~(\ref{p0}), we conclude that $ \Delta S_{1}^\textrm{re}(\tau)=0$, provided that system II is initially in a statistical mixture of its energy eigenstates. The heat flux is also zero for commuting interactions.
  Yet, if system I possesses some initial coherence, the dynamics would correlate the two partitions and the entropy production $\Delta S_{1}^\textrm{ir}(\tau) $ would exhibit oscillations. 
\begin{figure}
\includegraphics{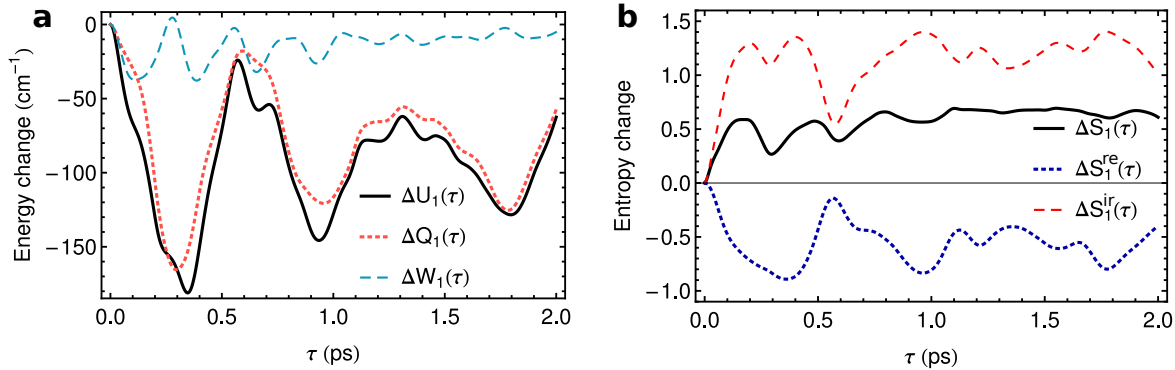}
\caption{(\textbf{a}) Heat and work contributions to the total energy change $\Delta U_1(\tau)$ of a two level system as it interacts with a single harmonic mode.
(\textbf{b}) Reversible and irreversible contributions to the change in the von Neumann entropy of the two level system.
Parameters are $T=300$~K,
$\omega_0 = 150~\textrm{cm}^{-1}$,
$V =  50~\textrm{cm}^{-1}$,
$\nu= 180~\textrm{cm}^{-1}$, and
$g= 50~\textrm{cm}^{-1}$.}
\label{fig:spinboson}
\end{figure}

To explore further possible relations between $ \Delta S_{1}^\textrm{re}(\tau) $ and $\Delta Q_1(\tau)$, we consider the non-commuting case where a two-level system interacts with a single harmonic mode. This 
spin-boson type Hamiltonian is given by
$H = (\frac{\omega_0}{2}\sigma_z +V\sigma_x) + \nu a^\dag a+ g\sigma_z (a+a^\dag)$, with bias $\omega_0$, tunnelling amplitude $V$, mode frequency $\nu$ and system-mode coupling strength $g$.
The spin is initially assumed to be in the excited state while the oscillator is initially in a thermal state at a temperature $T$. For these initial conditions the reversible entropy change quantifies the energy change of system II and becomes 
$\Delta S_{1}^{\textrm{re}} (\tau) =-\beta_2 (\langle H_2 \rangle_\tau-\langle H_2 \rangle_0)$
with $\beta_2=1/T$. Notice that this energy change for system II is not trivially related to the heat flux into system I.
Fig.~\ref{fig:spinboson}a displays the heat and work contributions to energy changes of the spin (parameters provided in the caption of Fig. \ref{fig:spinboson}), indicating there is predominantly a negative heat flux into the spin i.e. the system is ``cooled".
The reversible and irreversible contributions to entropy change of the spin are displayed in Fig. \ref{fig:spinboson}b. $\Delta S_1^\textrm{re}(\tau)$ is negative indicating that as the spin and oscillator become correlated, the latter evolves towards a non-thermal state with an average energy larger than its initial energy. Although $\Delta S_1^\textrm{re}(\tau)$ resembles some features of  $\Delta Q_1(\tau)$, such as its negativity and the position of some of the extrema, a proportionality relation of the form of Eq.~(8) in Ref.~\cite{Esposito:2010} does not follow straightforwardly.
We conclude this section by noting that although entropy production is positive its derivative can be negative. This is a direct consequence of the non-Markovian nature of the reduced dynamics of system I, and positivity of the rate is recovered in the Markov limit \cite{Esposito:2010}.  

 In the section, we have demonstrated that the energy flux formalism satisfies the second law of thermodynamics. By partitioning entropy changes as proposed in Ref. \cite{Esposito:2010}, the positivity of entropy production within the heat flux formalism is demonstrated. The direct relation between heat and entropy proposed in Ref. \cite{Weimer:2008} is incorrect in general.
 
 \section{Conclusions}
In this article we have considered internal exchange of energy in bipartite quantum systems and have investigated the consequences of the partitioning introduced in Ref. \cite{Weimer:2008} for different commutations of the interaction and the bare Hamiltonians.
A generalization of the formalism to Markovian open quantum systems was also presented and the consistency of the formalism with the second law of thermodynamics was discussed.  

From a classical standpoint exchange of heat and work have distinct physical origins; work is exchanged due to the action of an external drive. Heat, on the other hand, is exchanged with a thermal bath.  The present formalism connects these two scenarios and states how each situation arises in a particular limit of the general bipartite system. 

Optomechanical implementations provide the ideal platform for the investigation of energy transfer between quantum mechanical systems \cite{Verhagen:2012} and the prototype spin-oscillator models can readily be implemented in such a setting. Nano-scale thermoelectric devices comprised of quantum dots interacting with thermal reservoirs are alternative systems in which the interplay of quantum mechanical and thermodynamic principles can be investigated \cite{Sanchez:2011, Hartmann:2015}.  
There are, however, still issues associated with the measurement of the aforementioned fluxes and the practical applicability of the energy partitioning on the entropy fluxes requires further investigation.  

\section{Acknowledgements}
We acknowledge funding from the Engineering and Physical Sciences Research Council (EPSRC UK), Grant EP/G005222/1 and from the EU FP7 Project PAPETS (GA 323901).

\section{Bibliography}

\bibliographystyle{unsrt.bst}
\bibliography{main}

\section{Appendix A}  In this appendix we demonstrate that the reduced dynamics under a fully commuting interaction can be non-unitary. We consider the Hamiltonian $H = H_1 + H_2 + H_{12}$ where $[H_1, H_{12}] = [H_2, H_{12}] = 0$. The interaction Hamiltonian can be written as $H = A_1 \otimes B_2$ where $A_1$ acts on partition I and $B_2$ on partition II. $H_1$ and $A_1$ can be expanded in the energy eigenbasis of system I. That is, 
\begin{equation}
 H_1 = \sum_ i \epsilon_i  | \epsilon_ i \rangle \langle \epsilon_i  |, 
 \label{66}
\end{equation}
\begin{equation}
 A_1 = \sum_ i a_i  | \epsilon_ i \rangle \langle \epsilon_i  |, 
\end{equation}
where $\{ |\epsilon_i \rangle  \} $ are the eigenstates of system I with eigenenergies $\{ \epsilon_i\}$, and $\{ a_i \}$ are a set of coefficients. Similarly, we expand $H_2$ and $B_2$ in the eigenbasis of $H_2$
\begin{equation}
 H_2 = \sum_ i E_k  | E_ k \rangle \langle E_k  |,
\end{equation}
\begin{equation}
 B_2 = \sum_ k b_k  | E_ k  \rangle \langle E_k  |,
 \label{69}
\end{equation}
where $\{ | E_k \rangle \}$ are the eigenstates of system II with eigenenergies $\{ E_k \}$ and $\{ b_k \}$ are a set of coefficients.  
The reduced dynamics of partition I is given by $\rho_1 (t) = \mbox{Tr}_2 \big\{   \exp(-iHt)  \rho(0)  \exp{(iHt)} \big\}$. We expand the Hamiltonians and arrive at the following final expression
\begin{eqnarray}
 \rho_1 (t)  =   \sum_{ijk}   e^{-i( \epsilon_{i} - \epsilon_j) t} e^{-i (a_i - a_j ) b_k t}
| \epsilon_i \rangle \langle \epsilon_i | \rho_1 (0) | \epsilon_j \rangle 
\langle \epsilon_j | 
\otimes
\langle E_k |
 \rho_2 (0) 
 | E_k \rangle. 
\end{eqnarray}
 Unless partition II starts in an energy eigenstate, the dynamics of coherences $\langle \epsilon_i | \rho_1 (t) | \epsilon_j \rangle$ are non-unitary.  We therefore conclude that $\rho_1 (t)$ exhibits non-unitary dynamics provided that system II is not in an energy eigenstate at $t=0$, and system I possesses some initial coherence. Non-unitary dynamics give rise to a time-dependent von Neumann entropy for system I. 

\end{document}